\documentstyle[twocolumn,aps,prl]{revtex}
\include{epsf}
\epsfverbosetrue
\begin{document}
\twocolumn[\hsize\textwidth\columnwidth\hsize\csname@twocolumnfalse\endcsname

\draft

\title{Electron correlation effects and ferromagnetism in Iron}
\author{Pablo Pou, Fernando Flores, Jos\'e Ortega, Rub\'en P\'erez and Alfredo
Levy Yeyati}
\address{
Departamento de F\'{\i}sica Te\'orica de la Materia Condensada,
Universidad Aut\'onoma de Madrid, E-28049 Madrid, Spain}
\date{draft of \today}
\maketitle

\begin{abstract}
Electron correlation effects in Fe are analyzed
using a first principles LCAO-scheme. In our approach, we 
first use a local orbital
DFT-LDA solution to introduce a Hubbard Hamiltonian without
fitting parameters. In a second step, we
introduce a many-body solution to this Hamiltonian using a DMF
approximation. 
Our analysis shows that
magnetism in Fe is an effect associated with the first atomic
Hund's rule. Moreover, we also find important correlation
effects in the Fe-spin polarized DOS. 
The photoemision spectra is explained using a value of $U^{\mathrm{eff}}$ as large as 4
eV, provided the satellite peaks appearing around 3-5 eV below
the Fermi energy are interpreted appropriately.
\end{abstract}


]

\narrowtext

The electronic properties of ferromagnetic metals are still a
subject of controversy \cite{lichtenstein98,gutierrez,katsnelson}.
Although DFT-LDA calculations yield the correct magnetization for
the itinerant-electron ferromagnets Fe, Co and Ni, the origin of
ferromagnetism in these metals and the role of electron
correlations are not completely well-understood ({\it e.g.} see
\cite{kotliar}); in particular, the 
relative importance of the local Coulomb interaction for $d$ orbitals, $U^{\mathrm{eff}}$,
versus intraatomic exchange (first Hund's rule) is not completely
established. Even from an experimental point of view, there is
a lack of agreement on the existence or not of a satellite peak in
the photoemission spectrum of iron around 5 eV below the Fermi
energy \cite{gutierrez}.

In the conventional view of itinerant ferromagnetism \cite{ibach},
spin polarization is determined by the Stoner parameter, $I$, that defines
the energy of the atomic $d$-orbitals as $(E_d-I n_d)$, $n_d$
being the occupation number of the orbital under consideration. In
the case of Fe, DFT-LDA calculations yield a value $I = 3.9$ eV
\cite{papa}, and
a {\it surprisingly large} value of $U^{\mathrm{eff}}$ $\sim$
4--6 eV \cite{anisimov91,steiner} for
the effective Coulomb interaction between the $d$-electrons.
Since the atomic-like
properties of the $d$-states are of crucial importance for the
magnetic properties of these materials, LCAO methods
provide the appropriate conceptual framework to understand those
properties and to analyze the role of electron correlations.
In LCAO-theories of ferromagnetism, including Hubbard
Hamiltonians, $I$ is written as $(\widetilde{U}^{\mathrm{eff}}+4J^{x})$
\cite{majlis},
where $J^x$ defines the screened intrasite exchange interaction
between the atomic $d$-electrons having the same spin; it is
commonly accepted that $J^x$ practically coincides with its atomic
value \cite{anisimov91}. In the case of Fe, $J^x$ = 0.83 eV and,
therefore, we should take $\widetilde{U}^{\mathrm{eff}} \sim 0.6$ eV to
recover the value $I = 3.9$ eV that corresponds to the correct
magnetization. This result suggests the presence of dramatic
electron correlation effects in Fe, which would be responsible for the
renormalization of $U^{\mathrm{eff}}$ from  $\sim$ 5 eV to $\widetilde{U}^{\mathrm{eff}} \sim$
0.6 eV. On the other hand, the value of $U^{\mathrm{eff}}$ inferred from
the photoemission spectra \cite{turner,eastman,santoni} by
identifying photoemission peaks with quasi--particle peaks  yields
$U^{\mathrm{eff}} \approx$ 2 eV \cite{lichtenstein98,steiner}. This result
seems to indicate that electron correlation effects for Fe are not
strong, in contradiction with the previous analysis.

The purpose of this paper is to show that these apparent
contradictions disappear once electron correlation effects\cite{Stoddart71} are
properly analyzed using a first--principles LCAO--scheme. In our
approach, reminiscent of LDA+U \cite{anisimov97}, we first
formulate a Local Density solution for a generalized Hubbard
Hamiltonian.
This LD--solution provides the link between the generalized Hubbard
Hamiltonian and local orbital DFT-LDA methods, allowing us to
obtain that Hamiltonian from first principles,
without having to introduce fitting parameters. In a second step,
we introduce  a many-body solution for the Hubbard Hamiltonian
using a Dynamical Mean Field (DMF) approximation \cite{georges}:
in this way we analyze the spin-polarized electron density of
states for Fe and compare it with the experimental evidence
\cite{turner,eastman,santoni}. From our analysis, we obtain two
different results. First, using our LD-solution for the Hubbard
Hamiltonian, we show that electron correlation effects screen
strongly the effective Coulomb interaction
contributing to the Stoner parameter: in this sense,
$\widetilde{U}^{\mathrm{eff}}$ is not larger than 0.6-0.7 eV. We find,
however, that the effective interaction appearing in the Hubbard
Hamiltonian is around 4 eV, in reasonable agreement with other
first-principles calculations \cite{anisimov91,steiner}; using
this value and the many-body techniques mentioned above, we also
find that the spin-polarized DOS for Fe is in good agreement with
the photoemission data, provided we interpret appropriately the
satellite peaks appearing in the spectrum around 3-5 eV below the
Fermi energy \cite{steiner}.

Our starting point is the generalized Hubbard Hamiltonian:
\begin{eqnarray} \label{eqn:ham1}
& & \hat{H} = \hat{H}^{OE} + \frac{1}{2}\sum_{i,\alpha\sigma \neq \beta
\sigma'}U_{i}\hat{n}_{i\alpha\sigma}
\hat{n}_{i\beta\sigma'} \nonumber \\
& & - \frac{1}{2}\sum_{i,\alpha\sigma \neq \beta
\sigma}J^{x}_{i}\hat{n}_{i\alpha\sigma}
\hat{n}_{i\beta\sigma}
+\frac{1}{2} \sum_{i \neq
j}^{\alpha\sigma,\beta\sigma'} J_{i\alpha,j
\beta}\hat{n}_{i\alpha\sigma} \hat{n}_{j\beta\sigma'};
\end{eqnarray}
where $\hat{H}^{OE}$ defines a one-electron contribution, and
$U_i$ and $J_{i\alpha, j\beta}$ the intrasite and intersite
coulomb interactions between different orbitals $\phi_{i\alpha}$
and $\phi_{j\beta}$ (for the sake of simplicity, $U_i$ is an
average of the different interactions inside the $i$-site); we
also introduce the intrasite exchange coulomb interaction,
$J^{x}_{i}$, associated with the first atomic Hund's rule. Eqn.
\ref{eqn:ham1} has been written in an orthogonal local basis,
$\phi_{i\alpha}$, defined by the Lowdin's transformation
$\phi_{i\alpha} = \sum_{j\beta} (S^{-1/2})_{i\alpha,j\beta}
\psi_{j\beta} $, $\psi_{j\beta}$ being the local basis used in the
DFT-LDA calculation from which we obtain Hamiltonian
\ref{eqn:ham1}, as explained below ($S_{i\alpha,j\beta}$ is the
overlap between orbitals $\psi_{i\alpha}$ and $\psi_{j\beta}$).

The LD-solution of Hamiltonian \ref{eqn:ham1} is obtained by
introducing the kinetic and many-body energies of the system as a
function of the orbital occupancies, $n_{i\alpha\sigma}$
\cite{pou}. This implies that the total energy is a function of
those numbers, $n_{i\alpha\sigma}$, that play the role of the
electron density, $\rho(\bar{r})$, in  the conventional
DFT-approach. Then, we can write the following eqn.:

\begin{equation} \label{eqn:energy}
E[\{n_{i\alpha \sigma}\}]=
T[\{n_{i \alpha \sigma}\}] + E^{H}[\{n_{i\alpha \sigma}\}]
+ E^{XC}[\{n_{i\alpha \sigma}\}],
\end{equation}

where  $T=\langle \Psi_0 | \hat{H}^{OE} | \Psi_0 \rangle$,
$\Psi_0$ being the ground state of the total LD Hamiltonian;
$E^{H}$ is the hartree energy
and $E^{XC}$ is the exchange-correlation energy associated with
Hamiltonian \ref{eqn:ham1}. On the other hand \cite{pou}:
\begin{eqnarray}\label{eqn:energy-x}
E^{X}[\{n_{i\alpha\sigma}\}] = &-& \frac{1}{2}\sum_{i,\alpha\sigma
\neq \beta \sigma}J^{x}_{i}n_{i\alpha\sigma} n_{i\beta\sigma}
 \nonumber \\
 &-& \frac{1}{2}\sum_{i\alpha\sigma}
J_{i}n_{i\alpha\sigma} (1-n_{i\alpha\sigma}),
\end{eqnarray}
an eqn. that yields the exchange energy as the sum of an intrasite
contribution and of the intersite interaction between the electron
charge, $n_{i\alpha\sigma}$, and its hole,
$(1-n_{i\alpha\sigma})$. In this eqn., $J_{i}$ is practically the
coulomb interaction between charges located in n.n. atoms. Because
of the crystal symmetry , we assume that no exchange hole appears
in the same atom where the electron is located. On the other hand,
we have also shown \cite{pou} that the correlation energy is given
by:
\begin{eqnarray}\label{eqn:energy-c}
E^{C}[\{n_{i\alpha\sigma}\}] = -\frac{1}{2}\sum_{i\alpha\sigma}
f_{i} (U_i - J_i) n_{i\alpha\sigma} (1-n_{i\alpha\sigma})
\end{eqnarray}
where $(U_i-J_i)$ is an effective intrasite coulomb interaction
between $i$-site orbitals, and $f_i$ a quantity ranging between 0
and 1 depending on the importance of the intrasite correlation
effects  ($f_i$ is 1 for large values of $(U_i-J_i)$). Eqns.
\ref{eqn:energy}, \ref{eqn:energy-x} and \ref{eqn:energy-c} allow
us to substitute Hamiltonian \ref{eqn:ham1} for an effective
Hamiltonian where, instead of the many-body terms, we introduce
the local potentials $(V^{H}_{i\alpha\sigma}$ and
$V^{XC}_{i\alpha\sigma})$ given by:

\begin{eqnarray} \label{eqn:vh}
V^{H}_{i\alpha\sigma} &=& \frac{\partial E^{H}[\{
n_{i\alpha\sigma} \}]} {\partial n_{i\alpha\sigma}} =  \nonumber
\\ &=& \sum_{\beta \sigma' \neq \alpha \sigma} U_i n_{i\beta
\sigma'} + \sum_{j\beta \sigma' (j \neq i)} J_{i\alpha, j\beta}
n_{j\beta \sigma'}
\end{eqnarray}

\begin{eqnarray}\label{eqn:vxc}
V^{XC}_{i\alpha\sigma} &=& \frac{\partial
E^{XC}[\{n_{i\alpha\sigma}\}]} {\partial n_{i\alpha\sigma}}  = -
\sum_{\beta \neq \alpha} J^{x}_i n_{i\beta \sigma} \nonumber \\ &
& - J_{i}(\frac{1}{2}-n_{i\alpha\sigma})
  - f_i (U_i - J_i) (\frac{1}{2}-n_{i\alpha\sigma}),
\end{eqnarray}
where $f_i$ has been assumed to be constant.

This is the main result of our LD-analysis and shows how to reduce
the generalized Hubbard Hamiltonian, eqn. \ref{eqn:ham1}, to an
effective one-electron Hamiltonian, taking into account all the
many-body contributions. Conversely, we can use this equivalence
to go from a LD-solution to a generalized Hubbard Hamiltonian:
Assume we solve the conventional DFT-LDA eqns for a given crystal
(say, paramagnetic Fe) using a local orbital basis (as done in
Fireball \cite{fireball} or Siesta \cite{siesta} codes); then, we
can substract from the one-electron levels associated with the
orthogonalized orbitals, $\phi_{i\alpha}$, the potentials given by
eqns. \ref{eqn:vh} and \ref{eqn:vxc}. This difference defines
Hamiltonian $H^{OE}$ in eqn. \ref{eqn:ham1}, and allows us to
introduce the Hubbard Hamiltonian by means of the interactions
$U$, $J$ and $J^x$. Notice that in this approach we have to
calculate these interactions using the orthogonalized orbitals,
$\phi_{i\alpha}$. In our actual calculations, we have employed the
Fireball code for paramagnetic Fe and used the corresponding local
orbital basis.

   Ferromagnetic Fe has been analyzed in our LD-approach
by looking for a magnetic solution where some charge is  transferred  between
spins up and down. This implies that selfconsistent potentials,
$V_{i\alpha\uparrow}$ and $V_{i\alpha\downarrow}$,
should appear for different spins, in such a way that:
\begin{eqnarray}
V_{i\alpha\uparrow} &=& U_i (N_i-n_{i\alpha\uparrow}) +
\sum_{j\beta \sigma' (j \neq i)} J_{i\alpha, j\beta} n_{j\beta
\sigma'} \nonumber
\\ & & - J^{x}_i (N_{i\uparrow}- n_{i\alpha\uparrow})
- J_i (\frac{1}{2} - n_{i\alpha\uparrow}) \nonumber
\\ & & - f_i (U_i -J_i)
(\frac{1}{2} - n_{i\alpha\uparrow})
\end{eqnarray}

where $N_i$ is the total charge in the $d$-orbitals for the
$i$-site, while $N_{i\uparrow}$ represents the total spin-up
charge. Notice how the terms contributing to $V_{i\alpha\uparrow}$
correspond to the Hartree, the intraatomic exchange, the
extraatomic exchange and the correlation contributions,
respectively. Due to the magnetic polarization, we find changes in
the  many-body potential w.r.t.  the paramagnetic solution. This
yields:

\begin{eqnarray}\label{eqn:deltav}
\delta V_{i\alpha\uparrow}   &=&  - (1-f_i) (U_i -J_i) \delta
n_{i\alpha\uparrow} - J^{x}_i (\delta N_{i\uparrow}- \delta
n_{i\alpha\uparrow})
\end{eqnarray}
where the total charge, $N_i$, has been assumed to be constant and
independent from the atomic magnetization. Eqn. \ref{eqn:deltav}
defines how the $i\alpha\uparrow$-level depends on the atomic
polarization, $\delta n_{i\alpha\uparrow}$. This quantity should
be obtained selfconsistently by means of a band-structure
calculation whereby the atomic charges, $\delta
n_{i\alpha\uparrow}$, are a function of $\delta
V_{i\alpha\uparrow}$. These two conditions yield $\delta
n_{i\alpha\uparrow}$ and the crystal magnetization. Eqn.
\ref{eqn:deltav} allows us, however, to calculate directly the
Stoner parameter, $I$, which we define as $| \delta
V_{i\alpha\uparrow} / \delta n_{i\alpha\uparrow} |$. Eqn.
\ref{eqn:deltav} yields the following result:
\begin{equation}\label{eqn:stoner}
I= (1-f_i ) (U_i-J_i) + 4 J^{x}_i
\end{equation}
assuming $ \delta N_{i\uparrow} = 5 \delta n_{i\alpha\uparrow}$,
as corresponds to $d$ orbitals. In our calculations for Fe, we
find $U_i=$ 14.7 eV (taking into account atomic relaxation),
$J_i=$ 6 eV and $J^{x}_i$= 0.83 eV.

A word of caution should be introduced here, because in our discussion we
have neglected an effect that leads to a further reduction in
the effective interaction between orbitals. This is associated
with the $sp$-band screening that has been shown by other authors
\cite{anisimov91} to reduce $(U_i-J_i)=U^{\mathrm{eff}}_i$ to values close
to 5 eV. In our calculations, performed introducing a Lindhard
dielectric function, we have found that $U^{\mathrm{eff}}_i$ is reduced to
4.0 eV. If we introduce this value in eqn. \ref{eqn:stoner} and
take $f_i=$0.83 (the value that corresponds to this reduced
interaction), we find $I$ = 4.0 eV, in good agreement with DFT-LDA
calculations. We have also analyzed how $I$ depends on $U^{\mathrm{eff}}_i$
by calculating $f_i$ for different intrasite coulomb interactions.
Our results show that, in the 2-5 eV range, $(1-f_i)U^{\mathrm{eff}}_i$ is
almost insensitive to the values of $U^{\mathrm{eff}}_i$. These results
show that the Stoner parameter is mainly controlled by $J^{x}_i$;
in other words, ferromagnetism in Fe is an effect mainly
associated with the intraatomic first Hund's rule.

Next, we calculate many-body effects introducing a local
selfenergy, $\Sigma_{i\alpha\sigma}(\omega)$, within the
DMF-approximation. This is a reasonable approximation considering
that correlation effects in Fe are associated with the intrasite
coulomb interaction between $d$ orbitals. As discussed in ref
\cite{pou}, $\Sigma_{i\alpha\sigma}(\omega)$ is calculated by an
appropriate interpolation between two limits: (1) first , we
calculate the atomic limit, assuming $U^{\mathrm{eff}}_i$ much larger than
the metal bandwidth;(2) second, we obtain the second order
selfenergy, $\Sigma^{(2)}_{i\alpha\sigma}(\omega)$, using as the
expansion parameter $U^{\mathrm{eff}}_i$; (3) finally, we calculate the
selfenergy interpolating between these two limits.

We should stress that, in this solution,
$\Sigma^{(2)}_{i\alpha\sigma}$ is calculated  using the local
density of states defined by the LD-solution discussed above.
Finally, we  replace $V^{c}_{i\alpha\sigma}$, in our effective
LD-Hamiltonian, by that selfenergy; then, we use a conventional
Green-function formalism to calculate the local density of states.
At this point, we should comment that consistency between the LD-
and the selfenergy formulations imposes the following Luttinger
sum rule: $V^{c}_{i\alpha\sigma} = \Sigma_{i\alpha\sigma} (E_F)$.
The factor $f_i$ in eqn \ref{eqn:energy-c} has been determined
from this eqn., this procedure satisfying the Luttinger condition
automatically. The price we have to pay is the introduction of a
selfconsistent loop in the calculation.

Fig. 1 shows our calculated LD- and many-body DOS for
ferromagnetic Fe. The comparison between these two density of
states shows that correlation effects are important for Fe: first,
we notice that the energy difference between the two maxima
appearing in the spin-up and spin-down DOS for the LD-solution is
reduced by almost a factor of two in the many-body case. This is a
typical band narrowing effect appearing around $E_F$ and
associated with a highly correlated electron gas. On the other
hand, we also find that the DOS-structure at energies far away
from $E_F$ is strongly modified by the many-body solution: in
particular, a new satellite structure appears around 5 eV below
$E_F$. We analyze more in detail these many-body effects by
considering the DOS for a particular $\bar{k}$-vector: we have
chosen the $P$-point, a case for which there are high quality
photoemission data taken along the (111) direction
\cite{eastman,santoni}.

\begin{figure}[htbp]

\vspace*{-0.3cm}

\hspace*{+0.00cm} \epsfxsize=8cm \epsfbox{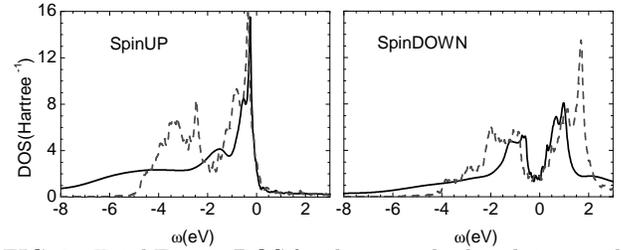}

\caption{Total Fe spin DOS for the many-body solution with $U^{\mathrm{eff}}$= 4 eV 
(continuous line) and for the DFT-LDA solution (dashed line). Notice the shift of the
peaks towards $E_F$ when correlation effects are introduced.}

\label{fig:1}
\end{figure}

\begin{figure}[htbp]

\vspace*{-0.6cm}

\hspace*{+0.00cm} \epsfxsize=8cm \epsfbox{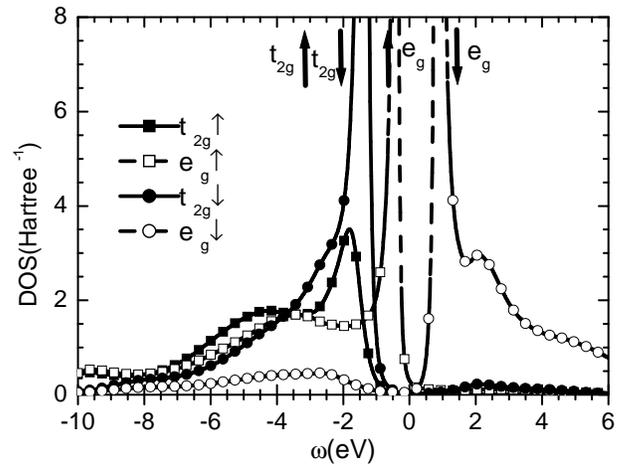}

\caption{Spin DOS, split into the $t_{2g}$ and $e_g$ components, 
at the $P$-point for Fe. 
The position of the DFT-LDA eigenvalues is indicated
by arrows. Notice that the t$_{2g}$(up) state (black squares)
has lost its quasiparticle character.}

\label{fig:2}
\end{figure}

\vspace*{-0.1cm}


Fig. 2 shows the electron DOS for this $\bar{k}$-vector split into
the $t_{2g}$ and the $e_{g}$-components, and
the energy levels corresponding to the DFT-LDA-solution for
this $\bar{k}$-point. In this figure we also find the effects
already discussed for fig. 1: the $e_{g}$(up) and $t_{2g}$(down)
levels located 0.6 eV and 2.1 eV below $E_F$  are shifted to
around 0.4 and 1.4 eV, respectively. In both cases, we also find
some satellites at higher binding energies due to many-body
effects. The other $t_{2g}$(up)-level located 3.1 eV below $E_F$
is completely smeared out by the selfenergy, giving rise to two
features: one is almost coinciding in energy with the
$t_{2g}$(down)-peak, the other one is a very broad peak located
around 4.5 eV below $E_F$. For energies above $E_F$, we find the
$e_{g}$(down)-level, located 1.4 eV above $E_F$,  shifted by
many-body effects to 0.9 eV. 
In fig. 3 we compare these results with
photoemission data \cite{eastman,santoni}, by considering the
appropriate weight that each state has in the photoemitted
electrons: this is done assuming the final state to be a plane
wave; in this way we find  that, for the $P$-point,
the $t_{2g}$-levels are reduced in
intensity by a factor of 3 w.r.t. the $e_{g}$-levels.
Comparing this DOS with the photoemitted spectra, we clearly see
that the two peaks below $E_F$ (at 0.4 and 1.4 eV) are related to
the $e_{g}$(up) and $t_{2g}$(down) states calculated in LD, while
the peak above $E_F$ (at 0.9 eV) corresponds to a $e_{g}$(down)
level: these results are in reasonable agreement with the
experimental data of refs. \cite{eastman,santoni}. More
importantly, we find that the very broad peak located around 3-4 eV below
$E_F$ cannot be related directly to the $t_{2g}$(up)-states found
in LD around 3.1 eV below $E_F$ \cite{eastman}. On the contrary,
our results clearly show that this peak is a satellite structure
created by many-body effects and appearing as the result of
combining the tail intensities of $e_{g}$ and $t_{2g}$ states (see
fig 2, and remember the factor 1/3 we have to introduce in the
weight of the $t_{2g}$-states).

\begin{figure}[htbp]

\vspace*{-0.3cm}

\hspace*{+0.00cm} \epsfxsize=8cm \epsfbox{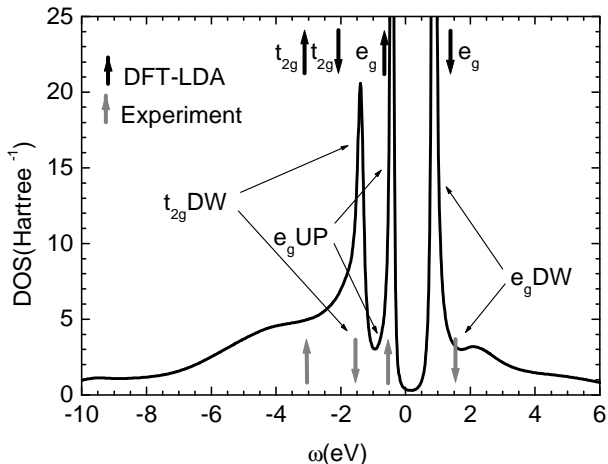}

\caption{Quasiparticle spectra (considering the appropriate weight for
each state, see text) at the $P$-point for Fe. Black
arrows indicate the eigenstates of the DFT-LDA Hamiltonian.
Experiments\protect\cite{eastman,santoni} show three well-defined peaks 
close to $E_F$ and a very broad feature around 3 eV below $E_F$ traditionally
interpreted as the $t_{2g}$(up) state (Gray arrows).}

\label{fig:3}
\end{figure}

\vspace*{-0.3cm}

This analysis clarifies several contradictory points about the
magnetism of iron. Our results show that we can find a reasonable
agreement between theory and photoemision data for $U^{\mathrm{eff}}$ as
large as 4 eV. The reason why a value of $U^{\mathrm{eff}}\simeq$ 2 eV has
been used in the interpretation of these data is the tendency to
identify the broad satellite peak, located around 3 eV below
$E_F$, at the $P$-point, with the quasiparticle level that in
DFT-LDA appears around 3.1 eV below $E_F$. Our results show,
however, that this quasiparticle level has lost its identity due
to correlation effects and that it has been modified into a
smeared DOS: this new DOS tends to create a satellite structure
that should be reinterpreted as due to many-body effects and not
as a quasiparticle level reminiscent of the DFT-LDA level. This
same effect is also responsible of the satellite peak we find in
the total DOS (see fig 1) around 5 eV below $E_F$.

In conclusion, we have studied the electronic properties of
ferromagnetic Fe using a first--principles LCAO--scheme to
analyze in detail the role of electron correlations. We find
that the correlation potential strongly screens the magnetic effects
commonly associated with a local Hubbard interaction:
magnetism in Fe is an effect associated with the first
atomic Hund's rule. Moreover, we also find
important correlation effects in the spin-polarized DOS.
In particular, our analysis shows that
the $t_{2g}$(up) level located 3.1 eV below $E_F$ have lost
their quasiparticle identity, due to many-body effects, and tend
to create a satellite structure that has been observed
experimentally.

This work has been partly funded by the spanish CICYT under contract
No. PB-97-0028. F. F. thanks N. H. March for an illuminating discussion.


\end{document}